\documentclass[prd,preprint,nofootinbib,showpacs]{revtex4}

\usepackage{amsmath,amssymb}
\usepackage{cancel}
\usepackage{graphicx}

\begin{document}

\preprint{CAFPE-143/10}
\preprint{UGFT-273/10}  

\title{Tau Custodian searches at the LHC}

\author{Francisco del Aguila} 
\author{Adri\'an Carmona} 
\author{Jos\'e Santiago}
\affiliation{
CAFPE and Departamento de F\'{\i}sica Te\'orica y del Cosmos,
\\
Universidad de Granada, E-18071 Granada, Spain
}


\begin{abstract}
The tau lepton can be more composite than naively expected 
in models of strong
electroweak symmetry breaking
with tri-bimaximal lepton mixing. New
leptonic resonances required by custodial symmetry, 
the tau custodians, 
can then be the first signal of this lepton flavor realization.
Tau custodians can be very light,
decaying almost exclusively into taus. 
The LHC reach for these new leptons is up to masses of 
240, 480 and 720 GeV 
for  $\sqrt{s}=14$ TeV and an integrated luminosity of
30, 300 and 3000 fb$^{-1}$, respectively.
Our analysis can be extended to any pair produced particles decaying
mostly into taus and Standard Model bosons.
\end{abstract}

\maketitle

\section{Introduction}

Custodial symmetry~\cite{Sikivie:1980hm} 
is a natural ingredient in models of strong
electroweak symmetry breaking (EWSB). The Standard Model (SM) fields
can be partly composite in these models~\cite{Kaplan:1991dc}, 
i.e. an admixture of
elementary and composite states, acquiring a mass from their
composite components. Thus, heavier fields are naturally more
composite, and also have a sizable mixing with composite states of the strong
sector with the same SM quantum numbers.
These composite states come, however, in full multiplets of
the custodial symmetry, the custodians, which can be relatively
 light and couple
strongly to the partly composite SM field. Then, it is natural
to expect, for instance, 
new light fermionic resonances with a large mixing with the
top~\cite{Carena:2006bn}.  

The leptonic sector can be similarly realized with an extra global
symmetry implying tri-bimaximal
mixing~\cite{Csaki:2008qq,delAguila:2010vg}.  
In this case, the tau
can be more composite than naively expected from its mass. Tau
custodians, the custodial symmetry partners of the composite
state mixing with the elementary tau, can then be relatively light, coupling
sizably only to the tau~\cite{delAguila:2010vg}. 
Moreover, these new resonances do not
disturb the very precisely measured properties of the tau lepton
because its coupling to the $Z$
boson is protected by a subgroup of the custodial symmetry
~\cite{Atre:2008iu,Albrecht:2009xr}.~\footnote{ 
In a similar way as originally proposed to protect the $Z b_L \bar{b}_L$
coupling~\cite{Agashe:2006at}.} 

In this letter we investigate the LHC reach for such new leptonic
resonances. They can be pair produced with electroweak (EW) strength 
through the exchange of a SM
gauge boson, decaying almost exclusively
into taus and a vector or scalar SM boson. This analysis is crucial
because signatures with taus in the final state are
typically deemed challenging and therefore not the first choice for
new physics searches. 
Such a signature could however very well be the first hint, and
maybe the only one for a while, of
the explicit realization of the lepton spectrum in models of 
strong EWSB.~\footnote{This is an interesting example in which the
  mechanism of neutrino mass generation, despite having a large
  suppression scale, has testable consequences at the LHC.} 
Pair production of these new resonances with the taus subsequently
decaying into leptons appears to be the cleanest, model independent
channel for these searches.
Assuming collinearity
and no other source of missing energy we can fully reconstruct the two
taus. Equality of the invariant mass of the two reconstructed new
leptons then allows to
reduce the background and reconstruct the custodian masses.

The outline of the paper is as follows. We review the main features
and signatures at the LHC of
the tau custodians in section~\ref{taucustodians:sec}.
The details of
the analysis and the results are given in section~\ref{analysis:sec}
and our conclusions in 
section~\ref{conclusions:sec}.

\section{New lepton doublets at the LHC\label{taucustodians:sec}}

The simplest realization of tau custodians with a protected $Z \tau
\bar{\tau}$ coupling consists of two vector-like lepton doublets with
hypercharges $-1/2$ and $-3/2$, 
\begin{equation}
L_{1L,R}^{(0)}=\begin{pmatrix} N^{(0)}_{L,R}
\\ E_{1L,R}^{(0)} \end{pmatrix}\sim (2)_{-\frac{1}{2}},
\quad
L_{2L,R}^{(0)}=\begin{pmatrix} E_{2L,R}^{(0)}
\\ Y_{L,R}^{(0)} \end{pmatrix}\sim (2)_{-\frac{3}{2}}, 
\end{equation}
respectively. The script $(0)$ indicates the current basis. 
The relevant part of the Yukawa and mass Lagrangian reads,
in the basis with diagonal charged
lepton Yukawa couplings, 
\begin{equation}
\mathcal{L}=-
\frac{m}{v} \bar{l}_L^{(0)}\varphi \tau^{(0)}_R
-\frac{m^\prime}{v} 
\Big[\bar{L}^{(0)}_{1L} \varphi + \bar{L}^{(0)}_{2L}
  \tilde{\varphi} \Big] \tau^{(0)}_R 
- M \Big[ \bar{L}^{(0)}_{1L} L^{(0)}_{1R}+ \bar{L}^{(0)}_{2L}
  L^{(0)}_{2R} \Big]+\mathrm{h.c.}
+\ldots,
\end{equation}
where the dots denote kinetic terms and other terms in the Lagrangian
not involving the new leptons. $\varphi$ is the SM Higgs doublet and
$\tilde{\varphi}= i \sigma^2 \varphi^\ast$, with $\sigma^2$ the second
Pauli matrix, $v\approx 174$ GeV is the Higgs vev, and
$l^{(0)}_L,\tau^{(0)}_R$ are the third generation SM leptons.
In the class of models we consider, the coupling to $e$, $\mu$ or any
right-handed neutrino is negligible.
After EWSB, the lepton mass matrix
\begin{equation}
\mathcal{M}=\begin{pmatrix}
m & 0 & 0 \\ m^\prime & M & 0 \\ m^\prime & 0 & M \end{pmatrix}
\end{equation}
is diagonalized with the usual
bi-unitary rotations, $U_L^\dagger \mathcal{M} U_R =
\mathcal{M}_{\mbox{diag}} = (m_\tau, m_{E_1},m_{E_2})$, 
which in our case take the very simple form
\begin{equation}
U_{L,R}=
\begin{pmatrix}
c_{L,R} & 0 & s_{L,R} \\
-\frac{s_{L,R}}{\sqrt{2}}
 & \frac{1}{\sqrt{2}} &  \frac{c_{L,R}}{\sqrt{2}} \\
-\frac{s_{L,R}}{\sqrt{2}}
 & -\frac{1}{\sqrt{2}} &  \frac{c_{L,R}}{\sqrt{2}} 
\end{pmatrix},
\end{equation}
where $s_{L,R} \equiv \sin (\theta_{L,R})$, $c_{L,R} \equiv \cos
(\theta_{L,R})$. 
All relevant physics can be parameterized in terms of $m$, $m^\prime$
and $M$.  However, it is simpler to use as alternative parameters
$m_\tau$, $s_R$ and $M$, where the latter two fully describe the
model, with
the left-handed mixing parameter
\begin{equation}
s_L= s_R \frac{m_\tau}{M}.
\end{equation}
In particular, assuming $M\geq 100$ GeV
 we have $s_L \leq 0.018$, $c_L\geq 0.9998$. (Thus 
$s_L\approx 0$, $c_L\approx 1$ is an excellent approximation.)
The resulting physical spectrum consists of three
degenerate leptons with mass $M$ and charges $0$, $-1$ and $-2$,
respectively 
\begin{equation}
m_N=m_{E_1}=m_Y=M,
\end{equation}
and a heavier charge $-1$ lepton with mass
\begin{equation}
m_{E_2}=\frac{M}{c_R}\sqrt{1-s_R^2\frac{m_\tau^2}{M^2}}.
\end{equation}

In the physical basis the lepton couplings
to the SM gauge bosons and to the Higgs can be written without
loss of generality
\begin{eqnarray}
\mathcal{L}^Z&=& \frac{g}{2c_W} \bar{\psi}^i_Q \gamma^\mu \Big[
X^{QL}_{ij} P_L +X^{QR}_{ij} P_R - 2 s_W^2 Q \delta_{ij} \Big]
\psi^j_Q Z_\mu, \\
\mathcal{L}^W&=& \frac{g}{\sqrt{2}} \bar{\psi}^i_Q \gamma^\mu \Big[
V^{QL}_{ij} P_L +V^{QR}_{ij} P_R \Big]
\psi^j_{(Q-1)} W^+_\mu + \mathrm{h.c.}, \\
\mathcal{L}^H&=& -\frac{H}{\sqrt{2}} \bar{\psi}^i_Q Y^{Q}_{ij} P_R \psi^j_Q +
\mathrm{h.c.},
\end{eqnarray}
where $Q$ runs over the electric charges in the spectrum ($-2, -1, 0$)
and $P_{LR}=(1\mp \gamma^5)/2$ are the chirality
projectors. 
In our case, the neutral gauge couplings read
\begin{eqnarray}
X^{(-1)}_L &=&
\begin{pmatrix}
-c_L^2 & s_L & -s_L c_L \\
s_L & 0 & -c_L \\
-s_L c_L & - c_L & -s_L^2
\end{pmatrix},
\quad
X^{(-1)}_R =
\begin{pmatrix}
0 & s_R & 0 \\
s_R & 0 & -c_R \\
0 & -c_R & 0 
\end{pmatrix},
\\
X^{(0)}_{L}&=& \begin{pmatrix} 1 & 0 \\ 0 & 1 \end{pmatrix},
\quad
X^{(0)}_{R}= \begin{pmatrix} 0 & 0 \\ 0 & 1 \end{pmatrix},
\quad
X^{(-2)}_{L}= X^{(-2)}_R=-1;
\end{eqnarray}
and the charged ones
\begin{eqnarray}
V^{(0)}_L &=&
\begin{pmatrix}
c_L U^{PMNS}_{33} & 0 & s_L  U^{PMNS}_{33}\\
-\frac{s_L}{\sqrt{2}} & \frac{1}{\sqrt{2}} & \frac{c_L}{\sqrt{2}}
\end{pmatrix},
\quad
V^{(0)}_R =
\begin{pmatrix}
0 & 0 & 0 \\
-\frac{s_R}{\sqrt{2}} & \frac{1}{\sqrt{2}} & \frac{c_R}{\sqrt{2}}
\end{pmatrix},
\\
V^{(-1)}_L &=&
\begin{pmatrix}
-\frac{s_L}{\sqrt{2}} & - \frac{1}{\sqrt{2}} & \frac{c_L}{\sqrt{2}}
\end{pmatrix}^\mathrm{T},
\quad
V^{(-1)}_R =
\begin{pmatrix}
-\frac{s_R}{\sqrt{2}} & - \frac{1}{\sqrt{2}} & \frac{c_R}{\sqrt{2}}
\end{pmatrix}^\mathrm{T},
\end{eqnarray}
where $U^{PMNS}_{33}$ is the corresponding entry of the PMNS
matrix~\cite{Pontecorvo:1957cp}. 
Finally, the corresponding Yukawa couplings read
\begin{equation}
v Y^{(-1)} =
\begin{pmatrix}
c_R^2 m_\tau & 0 & s_R c_R m_\tau \\
0 & 0 & 0 \\
s_R c_L M 
& 0 & 
\frac{s_R^2}{c_R} c_L M 
\end{pmatrix}.
\end{equation}
An explicit example, including
numerical values for these couplings in the context of composite
Higgs models, can be found in~\cite{delAguila:2010vg}.
Note that EW single production of these states in association with a tau
lepton is proportional to $s_L\approx 0$ or 
$s_R$, and therefore very sensitive to the particular value of 
the latter.
Pair production, on the other hand, is proportional to
the electric charge, to $c_L\approx 1$ or to $c_R$, 
and then less sensitive to the precise
value of $s_R$ unless $s_R\gtrsim 0.5$.
The three leptons with mass $M$ always decay into a tau lepton and a SM
gauge boson
\begin{equation}
N \to \tau W^+, \quad 
E_1 \to \tau Z, \quad  
Y \to \tau W^-,
\end{equation}
whereas the heavier one always decays to a tau and a Higgs
\begin{equation}
E_2\to \tau H,
\end{equation}
provided $c_R \geq (1+m_W/M)^{-1}$. For smaller $c_R$ values the
corresponding decay
channels into another heavy lepton and a gauge or Higgs boson open up. This
is an exciting possibility, since it allows for a richer phenomenology
but requires a large mixing (for
instance, $s_R\geq 0.5$ for $M\approx 720$ GeV). Mixing angles
that large require a detailed analysis of indirect constraints to
assess the phenomenological viability of the model and we
defer it to a future publication. Hence, we restrict ourselves to
the case in which all new leptons only decay to tau leptons and a
SM scalar or vector boson. 

New leptons can be singly produced in association with a tau or
pair produced at the LHC.
Single production, which may be relevant for
the early LHC run $\mathcal{L}\sim$ 1 fb$^{-1}$ at $\sqrt{s}=7$ TeV, 
is very sensitive to the values of the couplings in the model, as just
stressed. 
The relatively light masses and large couplings that can be tested in
this early run not only require an analysis of current EW constraints
but a dedicated study of the LHC reach, which will be presented elsewhere.
Pair production, on the other hand, is EW and model independent to a large
extent. 
The two heavy leptons then decay into two taus
and two SM bosons, which in turn will result in ten fermions in the
final state.
We are in the best
position to beat the background if we consider fully leptonic tau decays.
Besides, we will require a $Z$ in the final state decaying into
leptons for the same reason. 
Due to the relatively large mass of the heavy leptons, the two taus
are largely boosted and therefore their decay products highly
collimated. Assuming full collimation, we can completely reconstruct
the two taus despite having four neutrinos in the final state
if there is no further source of missing energy.
Thus, we consider the following channels
\begin{eqnarray}
&&pp \to \bar{E}_1 E_1 \to ZZ  \bar{\tau} \tau, 
\qquad
pp \to \bar{E}_1 Y \to Z W^-  \bar{\tau} \tau, 
\\
&&
pp \to \bar{E}_1 E_2 \to ZH  \bar{\tau} \tau, 
\qquad
pp \to \bar{E}_1 N \to Z W^+  \bar{\tau} \tau, 
\end{eqnarray}
together with the conjugated ones. 
The signature we are interested in is therefore
\begin{equation}
pp \to l^+ l^- l^{\prime +} l^{\prime\prime -} jj \cancel{E}_T,
\quad \mbox{with } l,l^\prime, l^{\prime\prime}=e,\mu.
\end{equation}
Even though we have to pay an important price due to the leptonic
branching ratios $\sim 0.6 \%$ [$\mathrm{BR}(Z\to l^+ l^-)\approx 6.6\%$,
$\mathrm{BR}(\tau \to l \cancel{E}_T)\approx
34\%$], the dramatic reduction of backgrounds overcomes this signal suppression.
Besides the multilepton final state, the full reconstruction of the
taus decaying leptonically and that the pair produced heavy leptons have
the same mass allows us to further reduce the
background down to an almost unobservable level.

\section{Analysis \label{analysis:sec}}

\begin{figure}[htb]
\begin{center}
\includegraphics[width=0.75\textwidth]{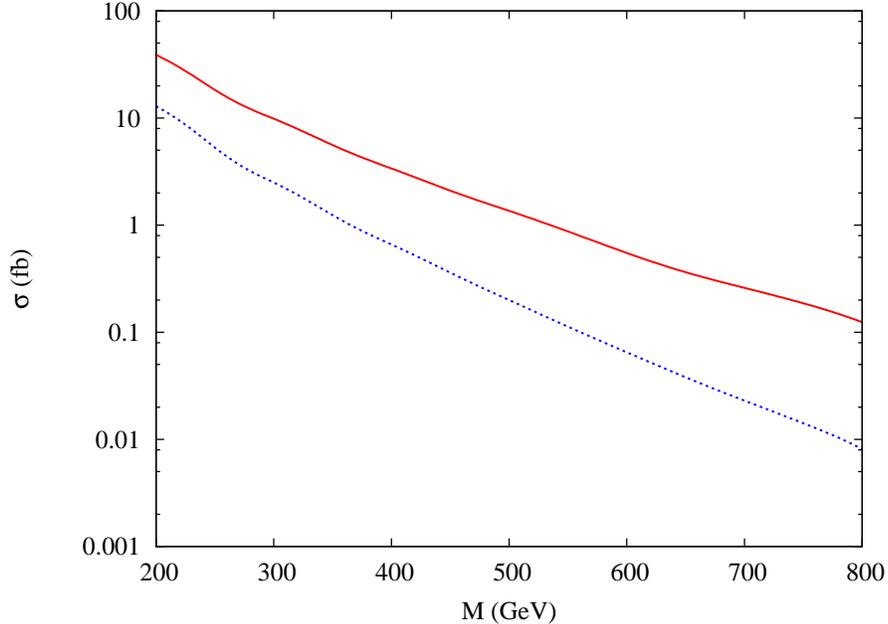}
\caption{ 
Heavy lepton pair production cross section 
(in fb) as a function of the heavy mass
$M$. The dotted (solid) line corresponds to $\sqrt{s}=7~(14)$ TeV. 
The cross section includes the leptonic $Z$ decay but not the tau
decays,
i.e. $pp\to l^+ l^- j j \tau^+ \tau^-$.
\label{fig:xsecsR01}
}
\end{center}
\end{figure}

As explained in the previous section, we consider pair
production of tau custodians for it is model independent. The
corresponding branching ratios, together with the
energy required to produce 
two heavy states makes the cross section too small to
have a significant number of events in the early LHC run. We thus
concentrate on the nominal energy 
$\sqrt{s}=14$ TeV.
The backgrounds we have considered are
\begin{eqnarray}
&&
Z t\bar{t}+n \mbox{ jets}, \quad \sigma = 39.6~\mathrm{fb}, 
\qquad 
Z b\bar{b}+n \mbox{ jets}, \quad \sigma = 5.85~\mathrm{pb}, 
\\
&&
ZZ +n \mbox{ jets}, \quad \sigma =  2.35~\mathrm{pb}, 
\qquad 
Z W +n \mbox{ jets}, \quad \sigma =  1.76~\mathrm{pb}. 
\\
&&
t\bar{t}+n \mbox{ jets}, \quad \sigma = 55~\mathrm{pb}, 
\qquad
ZWW+n \mbox{ jets}, \quad \sigma = 1.9~\mathrm{fb},
\end{eqnarray}
where $\sigma$ are the corresponding cross sections. One $Z$ 
in all channels and both tops in the $t\bar{t}$ channel 
have been required to decay leptonically and the cross section
reported includes the corresponding branching ratios and some minimal cuts.
In all cases we have generated up to $n=2$ jets at the partonic level 
with ALPGEN V2.13~\cite{ALPGEN}, and used the PGS4~\cite{PGS4} 
fast deterctor simulation after passing the events 
through PYTHIA~\cite{pythia} for hadronization and showering
(with the MLM matching algorithm). Our signal events are
generated with MADGRAPH/MADEVENT v4~\cite{Alwall:2007st} 
and taus are decayed with TAUOLA~\cite{Davidson:2010rw}. 
 In all
cases we have included initial and final state radiation but no
pile-up effects.
We show in Fig.~\ref{fig:xsecsR01} 
the signal production cross section, including the $Z$
leptonic branching ratio but not decaying the tau leptons, as a function of the heavy mass $M$
(and assuming a Higgs mass $m_H=120$ GeV).

In order to reduce the background we have implemented the following cuts
\begin{itemize}
\item \textbf{Basic cuts}. We require at least two positively 
 and two negatively charged isolated leptons (electrons or muons), two
 jets and missing energy with
\begin{eqnarray}
&&p_T(l)\geq 10~\mathrm{GeV},
\quad
p_T(j)\geq 20~\mathrm{GeV},
\quad
\cancel{E}_T\geq 20~\mathrm{GeV},
\nonumber \\
&&
|\eta_l|\leq 2.5,
\quad
|\eta_j|\leq 5,
\quad \Delta R_{jj} \geq 0.5,
\quad \Delta R_{jl} \geq 0.5.
\label{basic:cut}
\end{eqnarray}
We keep the hardest four leptons and two jets if their multiplicity is
larger.
\item \textbf{Leptons}. We require two same flavour, opposite charge leptons
  to reconstruct a $Z$, and the other two not to be back to back (so
  that the two taus can be reconstructed assuming collinearity),
\begin{equation}
|M_{l^+ l^-}-M_Z| \leq 10~\mathrm{GeV},
\quad
\cos (\phi_{l^{\prime +}l^{\prime\prime -}}) \geq -0.95,
\end{equation}
\item $\mathbf{M_{jj}}$. The two jets in our signal come from the
  decay of a SM boson. We therefore 
impose a cut on the invariant mass of
  the two jets
\begin{equation}
50~\mathrm{GeV} \leq M_{jj} \leq 150~\mathrm{GeV}.
\end{equation}
\item \textbf{$\tau$ reconstruction}. We use the two leptons not
  reconstructing the
  $Z$ and the transverse missing energy to infer the tau
  four-momenta~\cite{Rainwater:1998kj}.
First, we assume all momenta in the tau decays are aligned
\begin{eqnarray}
p^{l^{\prime +}}_i&=& x^+ p^{\tau^+}_i, \quad \cancel{p}^+_i = (1-x^+)
p^{\tau^+}_i,
\\
p^{l^{\prime\prime -}}_i&=& x^- p^{\tau^-}_i, \quad \cancel{p}^-_i = (1-x^-)
p^{\tau^-}_i,
\end{eqnarray}
where $i$ stands for the spatial components $x,y,z$ and
$\cancel{p}^\pm_i$ denotes the sum
of the momenta of the neutrinos coming from the $\tau^\pm$ decay.
$x^\pm$ are the fraction of $\tau^\pm$ momentum taken by $l^{\prime
  +},l^{\prime\prime-}$, respectively. They are fixed by momentum
conservation in the transverse plane 
\begin{eqnarray}
x^+ &=&
\frac{
p^{l^{\prime\prime -}}_y p^{l^{\prime +}}_x 
-p^{l^{\prime\prime -}}_x p^{l^{\prime +}}_y
}{
\cancel{p}_x p^{l^{\prime\prime -}}_y 
-\cancel{p}_y p^{l^{\prime\prime -}}_x 
+p^{l^{\prime\prime -}}_y p^{l^{\prime +}}_x 
-p^{l^{\prime\prime -}}_x p^{l^{\prime +}}_y
},
\\
x^- &=&
\frac{
p^{l^{\prime\prime -}}_y p^{l^{\prime +}}_x 
-p^{l^{\prime\prime -}}_x p^{l^{\prime +}}_y
}{
\cancel{p}_y p^{l^{\prime +}}_x 
-\cancel{p}_x p^{l^{\prime +}}_y 
+p^{l^{\prime\prime -}}_y p^{l^{\prime +}}_x 
-p^{l^{\prime\prime -}}_x p^{l^{\prime +}}_y
}.
\end{eqnarray} 
These lie between 0 and 1 if all transverse missing energy, measured
with infinite precision, comes from 
collinear tau decays. Thus, we require
$0\leq x^\pm \leq 1$ and use them to reconstruct the $\tau^\pm$
four-momenta 
\begin{eqnarray}
p^{\tau^+}_i&=&\frac{p^{l^{\prime +}}_i}{x^+}, \quad
p^{\tau^-}_i=\frac{p^{l^{\prime\prime -}}_i}{x^-}, \quad
i=x,y,z, 
\\
p^{\tau^\pm}_0 &=& \sqrt{m_\tau^2 + \sum_{i=x,y,z}
  (p^{\tau^\pm}_i)^2}.
\end{eqnarray}

\item \textbf{Pair production}. We require the two reconstructed heavy
  leptons to have the 
  same mass within 50 GeV,
\begin{equation}
|M_{L_1} - M_{L_2}|\leq 50~\mathrm{GeV},
\end{equation}
where $M_{L_i}$ corresponds to the invariant mass of $\tau^\pm$ and
either $l^+ l^-$ or $jj$. (We select the pairing giving the smaller
difference.) 

\item\textbf{Mass reconstruction}. Finally we require the invariant mass of
  the $\tau l^+ l^-$ pairing to peak around a
  test mass within 50 GeV. 
\begin{equation}
|M_{\tau l^+ l^-} - M_{L^\mathrm{test}}| \leq 50~\mathrm{GeV}.
\label{Mrec:cut}
\end{equation}
\end{itemize}

\begin{figure}[ht]
\begin{center}
\includegraphics[width=0.75\textwidth]{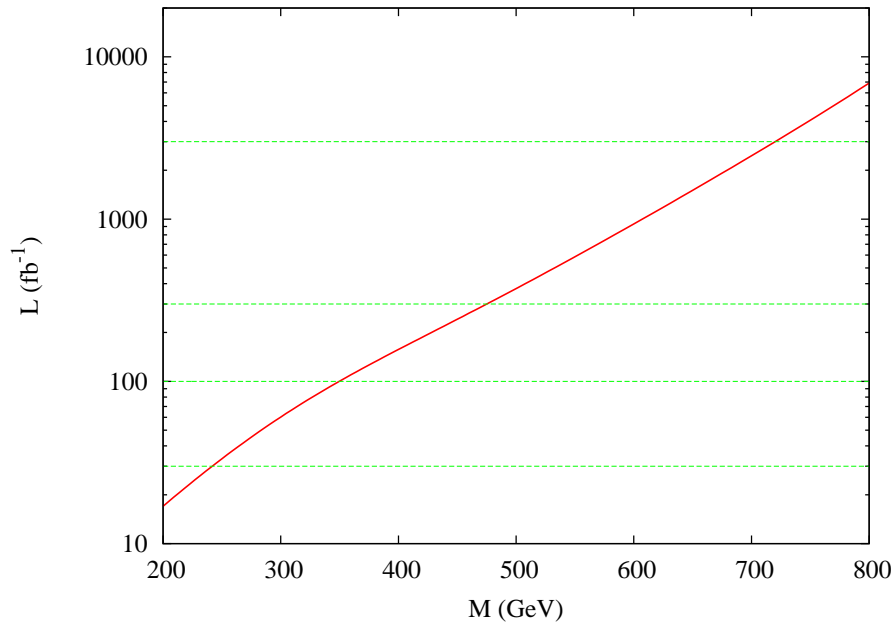}
\caption{ 
Luminosity required for a 5 $\sigma$ discovery at the LHC with
$\sqrt{s}=14$ TeV as a
function of the custodian mass $M$.
\label{fig:luminosity5sigma}
}
\end{center}
\end{figure}

We have applied the analysis described above to the signal, for
different values of the custodian mass $M$, and to the 
background. In order to estimate the statistical significance of the
result we use
\begin{equation}
S_{cL}=\sqrt{2\Big((s+b)\ln (1+s/b)-s\Big)},
\end{equation}
where $s$ and $b$ are the number of signal and background events,
respectively, after all cuts have been imposed~\cite{Ball:2007zza}. 
We require a minimum number of 3 signal events and $S_{cL}=5$ for
a $5\sigma$ discovery.
An example of the efficiency of each cut on the signal and on the main
backgrounds for two sample custodian masses $M=200$ GeV and
$M=400$ GeV is shown in
Table~\ref{tab:cuts}. 
\begin{table}[ht]
\begin{center}
\begin{tabular}{|c|c|c|c|c|}
\hline
14 TeV & $M=200$ GeV   & $M=400$ GeV  & $Z t\bar{t}$ & $ZZ$ \\
\hline
Basic & 0.85 & 0.14 &  0.49 & 0.44 \\
Leptons & 0.68 (81\%)  & 0.11 (77\%) & 0.41(84\%)& 0.41 (93\%)  \\
$M_{jj}$ & 0.49 (72\%)& 0.063 (59\%)& 0.15(37\%)&0.13 (31\%) \\
Tau rec. & 0.42 (86\%)& 0.057 (90\%) & 0.039 (26\%)&0.052(40\%) \\
Pair prod. & 0.39 (91\%)&0.045 (79\%)  & 0.017(44\%)&0.032 (61\%)  \\
Mass rec. & 0.37 (96\%)&0.041 (91\%) & 0.008 (48\%) \Big|
0.0016 (9\%) & 
0.016 (50\%) \Big | 0.0018 (6\%)  \\
\hline
\end{tabular}
\caption{Cross sections in fb (and corresponding efficiencies) 
after cuts for the signal
  and main backgrounds. The cuts are described in
  Eqs. (\ref{basic:cut}-\ref{Mrec:cut}). We show the results for 
two different
  values of the custodian masses $M=200$, $400$ GeV. The effect
  of the last cut on the background depends on the test mass as shown
  in the last row.
The required luminosity to have a 5 $\sigma$ discovery,
  with 3 or more events, being
$\mathcal{L}\approx  17$,
$170~\mathrm{fb}^{-1}$, respectively.
\label{tab:cuts}}
\end{center}
\end{table}
The required luminosity for a $5\sigma$
discovery is 17 and 170 fb$^ {-1}$, respectively.
The corresponding luminosity as a function of the custodian masses
is shown in 
Fig.~\ref{fig:luminosity5sigma}. The expected reach after 30, 300 and
3000 fb$^{-1}$ 
of integrated luminosity is $M\sim 240$, $480$ and $720$ GeV,
respectively, for a 5 $\sigma$ discovery.

\section{Conclusions \label{conclusions:sec}}

New light leptonic resonances related to the tau lepton through
custodial symmetry, tau custodians, 
can be a natural occurrence in models of strong
EWSB, if a global symmetry governs the lepton spectrum.
Thanks to the custodial symmetry, they can be light and strongly
coupled to the tau without conflict with EW precision
or flavour data. Pair production of tau custodians provides a clean, model
independent channel, that results in two taus and two gauge or Higgs
bosons. Requiring at least one $Z$ decaying into
electrons or muons, leptonic tau decays and no further source of 
missing energy,
we end up with a final state with four charged leptons (electrons or
muons), missing energy and two jets. The large number of leptons
allows for a very efficient reduction of the main backgrounds.
The relative large mass of the custodians results in highly boosted
taus with very collimated decay products. Assuming complete collimation,
we can fully reconstruct both taus, despite the presence of four
neutrinos in the final state. The requirement of pair production of
same mass objects then further enhances the signal, leading to a
discovery reach for tau custodians at the LHC with
$\sqrt{s}=14$ TeV of $M=240$, $480$ and $720$ GeV 
for a total integrated luminosity $\mathcal{L}=30$, $300$,
$3000~\mathrm{fb}^{-1}$, respectively.

This analysis is crucial in the context of models of strong
EWSB due to the difficulty of observing pair production of tau
custodians
because they only decay into taus, and this could 
be the very first experimental signature of the explicit
realization of the lepton spectrum in these models. 
It can be applied to any new particles
that are pair produced and decay predominantly into taus and gauge or
Higgs bosons. Hadronic tau decays could be also used to search for
these new resonances. A rough estimate indicates that they could give
a similar sensitivity. However, the a priori larger backgrounds and the
need of an efficient tau identification make a full real detector
simulation compulsory. What would be also welcome for the analysis
presented here.

\begin{acknowledgments}
It is a pleasure to thank J.A. Aguilar-Saavedra, J. Alwal, M. Herquet,
R. Pittau and especially M. Treccani for useful discussions.
This work has been partially supported by MICINN project FPA2006-05294
and Junta de Andaluc\'{\i}a projects FQM 101, FQM
03048. A.C. is partially supported by an FPU fellowship and J.S. by a
Ram\'on y Cajal contract. 
\end{acknowledgments}

\end{document}